\documentclass[showpacs,aps,prd]{revtex4}
\usepackage{graphicx} % 插图支持
\usepackage{epstopdf}
\usepackage[justification=raggedright]{caption}
\usepackage{amsmath}  % 数学公式
\usepackage{amssymb}  % 数学符号扩展
\usepackage{bm}       % 粗体数学符号
\usepackage{slashed}  % Feynman slash 记号
\usepackage{tikz} % 用于绘制图形
\usepackage{subcaption} % 用于子图排版
\usepackage{booktabs} % 专业表格线
\usepackage{titlesec}
\usepackage{nicefrac}
\usepackage{natbib}
\usepackage{romannum}
\usepackage[english]{babel}
\usepackage[UTF8]{ctex}
\usepackage[colorlinks=true, linkcolor=blue, citecolor=blue, urlcolor=blue]{hyperref}

\input epsf

\usepackage{pgfplots} % 用于绘制数据图

\begin{document}

\title{Fully heavy pentaquark states from QCD sum rules}
\author{Zheng-Lei Jing and Jian-Rong Zhang}
\affiliation{College of Science, National University of Defense Technology,Changsha 410073, Hunan, People's Republic of China}

\date{\today}

%%%%%%%%%%%%%%%%%%%%%%%%%%%%%%%%%%%%%%%%%%%%%%%%%%%%%%%%%%%%%%%%%%%%%
\begin{abstract}
In this work, we systematically investigate the mass spectra of fully heavy pentaquark states $QQQQ\bar{Q}$ within the framework of QCD sum rules. Employing three configurations of interpolating currents, we obtain the following mass predictions: for the $cccc\bar{c}$ states, the masses are determined to be $7.79^{+0.18}_{-0.17}$ GeV, $7.76^{+0.23}_{-0.18}$ GeV, $7.81^{+0.17}_{-0.18}$ GeV; while for the $bbbb\bar{b}$ states, the corresponding masses are calculated as $22.35\pm0.19$ GeV, $22.27\pm0.22$ GeV, $22.37^{+0.18}_{-0.20}$ GeV, respectively.
\end{abstract}
\maketitle

\section{INTRODUCTION}\label{sec1}
Hadron physics represents a pivotal frontier in contemporary particle physics research, primarily investigating bound states of quarks and gluons governed by strong interactions. 
Conventionally, hadronic states are categorized into two fundamental types: mesons and baryons. 
However, quantum chromodynamics (QCD) imposes no fundamental constraints limiting hadrons to conventional configurations. 
The theory naturally accommodates more complex structures, including multiquark systems  and gluonic bound states, which are collectively termed exotic hadrons.
Currently, the exploration of exotic hadron states stands as one of the most significant and active research domains within hadron physics.
One could see some recent reviews \cite{RNad1,RNad10,RNad2,RNad3,RNad4,RNad5,RNad6,RNad7,RNad11} and references therein.

The investigation of exotic hadron states gained significant promotion with the discovery of the $X(3872)$ resonance \cite{RN1}. 
Observed by the decay channel $B^{\pm}\rightarrow K^{\pm}\pi^{+}\pi^{-}J/\psi $, this state displayed characteristics that challenged conventional quark model predictions. 
Its unusual properties stimulated intense theoretical and experimental scrutiny of exotic hadronic matter, including observations of pentaquark states with the discovery of the $P_c(4380)^+$ and $P_c(4450)^+$ resonances by the LHCb Collaboration in the $\Lambda_b^0 \to J/\psi p K^-$ decay channel \cite{RN2}. 
Subsequent analysis with improved statistics in 2019 revealed a new state, $P_c(4312)^+$, while resolving the broad $P_c(4450)^+$ peak into two narrow structures, $P_c(4440)^+$ and $P_c(4457)^+$ \cite{RN3}.
The LHCb Collaboration has further reported two additional observations: the $P_{cs}(4459)^0$ state, identified in the $J/\psi\Lambda$ mass spectrum of $\Xi_{b}^{-} \rightarrow J/\psi \Lambda K^-$ decays \cite{RN4}, and the $P_{cs}(4338)^0$ state, found through amplitude analysis of $B^{-} \rightarrow J/\psi \Lambda \bar{p}$ decays \cite{RN5}. Moreover, the $P_c(4337)^+$ state observed by LHCb in $B_{s}^{0} \rightarrow J/\psi p \bar{p}$ decays has been considered to be a promising pentaquark state candidate \cite{RN6}.
In 2020, the LHCb Collaboration reported a narrow structure $X(6900)$ and proposed it to be a all-charmed tetraquark state \cite{RN7}, which has greatly stimulated the study of fully heavy multiquark states.

On investigations of fully heavy pentaquark states, researchers have employed a variety of theoretical methodologies \cite{RN8,RNzjr,RN9,RN10,RN11,RN12,RN13,RN14,RN15,RN16,RN17,RN18,RN19,RN20,RNad9}.
Concretely, studies have been conducted within  the chromomagnetic interaction model \cite{RN8}, chiral quark model \cite{RN10,RN14}, Lattice-QCD inspired quark model \cite{RN12}, MIT bag model \cite{RN13}, and nonrelativistic potential quark model \cite{RN15}. 
In addition, analyses of fully heavy pentaquark states have been carried out utilizing approaches such as the variational method within the constituent quark model \cite{RN11}, the diffusion Monte Carlo method \cite{RN17},  the Suzuki model \cite{RN19}, and the extended form of the Gursey-Radicati mass formula \cite{RN20}.

During the study of fully heavy pentaquark states, it is inevitably to confront the highly intricate nonperturbative problems inherent in QCD.
As one reliable method for evaluating nonperturbative effects, the QCD sum rule \cite{RN25,RN26} is firmly established on the basic QCD theory and has been successfully used in the study of hadronic states \cite{RNadd1,RNadd2,RNadd3,RNadd4,RNad8}.
By far, some works have made use of QCD sum rules to address the nonperturbative issues on fully heavy pentaquark states \cite{RNzjr,RN9,RN16,RN18}.
In Ref. \cite{RNzjr}, the author gained a mass of $7.41^{+0.27}_{-0.21}$ GeV for $cccc\bar{c}$ states and $21.60^{+0.73}_{-0.22}$ GeV for $bbbb\bar{b}$ states, suggesting that these states could be observed in the $\Omega_{QQQ} \eta_Q$ invariant mass spectrum.
Ref. \cite{RN9} focused on the diquark-diquark-antiquark type fully heavy pentaquark states with the spin-parity $J^{P}=\frac{1}{2}^{-}$, obtaining the mass of $cccc\bar{c}$ to be $7.93\pm 0.15$ GeV and that of $bbbb\bar{b}$ to be $23.91\pm0.15$ GeV. 
The author proposed that these states might be detectable in the \(J/\psi\Omega_{ccc}\) and \(\Upsilon\Omega_{bbb}\) invariant mass spectrum.
Ref. \cite{RN16} researched the full-charm and full-bottom pentaquark states with spin-parity \(\frac{3}{2}^{-}\), reporting masses of $7628\pm112$ MeV and $21982\pm 144$ MeV, respectively.
In Ref. \cite{RN18}, the authors analyzed the fully heavy pentaquark candidates with quark contents of $cccb\bar{c}$ and $bbbc\bar{b}$.
For the $P_{4cb}$ states, they obtained masses of $11388.30\pm107.79$ MeV and $11368.30\pm112.68$ MeV, while for $P_{4bc}$ states, the reported masses were $20998.30\pm121.52$ MeV and $20990.50\pm125.87$ MeV.

In this paper, we devote to extend our previous research \cite{RNzjr} and systematically investigate the fully heavy pentaquark states from QCD sum rules. 
The remainder of the paper is organized as follows: the derivation of the QCD sum rules for fully heavy pentaquark states is presented in section \ref{two}, and the numerical analysis and discussions are given in section \ref{three}. 
The final section comprises a concise summary.


\section{FULLY HEAVY PENTAQUARK STATES IN QCD SUM RULES}\label{two}
The interpolating current representing a fully heavy pentaquark state can be constructed as different configurations, such as the diquark-diquark-antiquark arrangement adopted in Ref. \cite{RN9}, or compositions of the baryon and meson currents.
In this work, we will employ the fully heavy pentaquark state current in the baryon-meson form.

Based on the interpolating current construction ways for heavy mesons \cite{RN36} and baryons \cite{RN37,RN38} in full QCD, we can formulate following forms of currents
\begin{equation}
	j_{\mu} = ( \epsilon_{abc} \, Q_{a}^{T} C \gamma_{\mu} \, Q_{b} \, Q_{c} ) ( Q_{e} \bar{Q}_{e} ), \label{0}
\end{equation}
\begin{equation}
	j_{\mu\rho} = ( \epsilon_{abc} \, Q_{a}^{T} C \gamma_{\mu} \, Q_{b} \, Q_{c} ) ( Q_{e} \gamma_{\rho} \bar{Q}_{e} ), \label{1}
\end{equation}
and
\begin{equation}
	j_{\mu\rho} = ( \epsilon_{abc} \, Q_{a}^{T} C \gamma_{\mu} \, Q_{b} \, Q_{c} ) ( Q_{e} \gamma_{\rho} \gamma_{5} \bar{Q}_{e} ) \label{2}
\end{equation}
for $QQQQ\bar{Q}$ pentaquark states.	
Here, $T$ denotes the transpose of a matrix, $C$ denotes the charge conjugate matrix, and $Q$ can be either a heavy charm or bottom quark. The subscripts $a$, $b$, $c$, and $e$ are colour indices.

In order to derive the QCD sum rule, one can construct the two-point correlator
\begin{equation}
	\Pi_{\mu\nu}(q^2) = i \int d^4x \, e^{iq \cdot x} \langle 0 | T \left[ j_\mu(x) \bar{j}_\nu(0) \right] | 0 \rangle 
\end{equation}
for the current in Eq. (\ref{0}).
The parameterization of $\Pi_{\mu\nu}(q^2)$ is as follows
\begin{equation}
	\Pi_{\mu\nu}(q^2) = -g_{\mu\nu} \left[\slashed{q} \Pi_1(q^2) + \Pi_2(q^2) \right] + \cdots.
\end{equation}

For the currents in Eq. (\ref{1}) and Eq. (\ref{2}), the two-point correlator 
\begin{equation}
	\Pi_{\mu\rho\nu\sigma}(q^2) = i \int d^4x \, e^{iq \cdot x} \langle 0 | T \left[ j_{\mu\rho}(x) \bar{j}_{\nu\sigma}(0) \right] | 0 \rangle
\end{equation}
can be parameterized as
\begin{equation}
	\Pi_{\mu\rho\nu\sigma}(q^2) = -g_{\mu\nu}g_{\rho\sigma} \left[\slashed{q} \Pi_1(q^2) + \Pi_2(q^2) \right] + \cdots.
\end{equation}

By matching the terms proportional to $-g_{\mu\nu}\slashed{q}$ or $-g_{\mu\nu}g_{\rho\sigma}\slashed{q}$ in the hadronic representation with the quark representation and after applying the Borel transform, one gets
\begin{equation}
	\lambda^2 e^{-M_H^2/M^2} = \int_{25m_Q^{2}}^{s_0} ds \, \rho \, e^{-s/M^2},  \label{3}
\end{equation}
where $M_H$ is the mass of the studied hadron and the spectral density is $\rho=\frac{1}{\pi}$Im$\Pi_1(s)$.
The hadron mass is obtained by taking the derivative of Eq. (\ref{3})  with respect to $-\frac{1}{M^2}$ and dividing by Eq. (\ref{3}) itself.
The expression is
\begin{equation}
	M_H = \sqrt{  \int_{25m_Q^{2}}^{s_0} \! ds \, \rho s \, e^{-s/M^2}  \Big/  \int_{25m_Q^{2}}^{s_0} \! ds \, \rho \, e^{-s/M^2}}. \label{4}
\end{equation}

In OPE calculations, one usually employs heavy quark propagators in momentum space \cite{RN36}. 
By extending the relevant techniques \cite{RN39,RN40,RN41,RN42,RN43,RN44,RN45} to fully heavy pentaquark systems as Ref. \cite{RNzjr}, the specific expression for the spectral density $\rho= \rho^{\text{pert}} + 	\rho^{\left\langle g^2 G^2\right\rangle } + \rho^{\left\langle g^3 G^3\right\rangle} $ in Eq. (\ref{4}) is calculated to be

\begin{widetext}
	\[
	\begin{split}
		\rho^{\text{pert}} = & -\frac{3}{5\times2^{14}\pi^8}
		\int_{\alpha_{min}}^{\alpha_{max}} \frac{d\alpha}{\alpha^3} 
		\int_{\beta_{min}}^{\beta_{max}} \frac{d\beta}{\beta^3} 
		\int_{\gamma_{min}}^{\gamma_{max}} \frac{d\gamma}{\gamma^3} 
		\int_{\xi_{min}}^{\xi_{max}} \frac{d\xi}{\xi^3} 
		\frac{\mathbf{h}^3(m_Q^2 - \mathbf{h}s)^3}{(1 - \alpha - \beta - \gamma - \xi)^3} 
		\Bigg\{ 
		8 \mathbf{h}^2 (m_Q^2 - \mathbf{h}s)^2\\
		& - \left[ 35 \mathbf{h}^3 s + 5 \mathbf{h} m_Q^2 (3\alpha \beta-2\gamma \xi) \right]  (m_Q^2 - \mathbf{h}s)  
		+ 20 \mathbf{h}^4 s^2 
		+ 20 \mathbf{h}^2 s m_Q^2(\alpha \beta - \gamma \xi) 
		-20 \alpha \beta \gamma \xi m_Q^4 
		\Bigg\},
	\end{split}
	\]
	\[
	\begin{split}
		\rho^{\left\langle g^2 G^2\right\rangle } =& -\frac{m_Q^2\left\langle g^2G^2\right\rangle }{2^{14}\pi^8} 
		\int_{\alpha_{\min}}^{\alpha_{\max}} \frac{d\alpha}{\alpha^3} 
		\int_{\beta_{\min}}^{\beta_{\max}} \frac{d\beta}{\beta^3} 
		\int_{\gamma_{\min}}^{\gamma_{\max}} \frac{d\gamma}{\gamma^3} 
		\int_{\xi_{\min}}^{\xi_{\max}} \frac{d\xi}{\xi^3} 
		\frac{\mathbf{h}^3}{(1-\alpha - \beta - \gamma - \xi)^3} 
		\Bigg\{ 
		\Big[(1-\alpha - \beta - \gamma - \xi)^3 \\
		& + 2\beta^3 + 2\xi^3\Big] 
		\Big[4\mathbf{h}^2(m_Q^2 - \mathbf{h}s)^2 
		- 7\mathbf{h}^3 s(m_Q^2 - \mathbf{h}s) 
		+ \mathbf{h}^4 s^2 
		- \alpha \beta \gamma \xi m_Q^4 
		- \mathbf{h} m_Q^2 (3\alpha \beta-2\gamma \xi)(m_Q^2 - \mathbf{h}s)  \\
    	&+ \mathbf{h}^2 s m_Q^2(\alpha \beta - \gamma \xi)\Big]
		+ 3\mathbf{h} (2\gamma \xi^3-3 \alpha \beta^3) (m_Q^2 - \mathbf{h}s)^2 
		- 6(\gamma \xi^3 - \alpha\beta^3) \mathbf{h}^2 s (m_Q^2 - \mathbf{h}s)\\ 
		&- 6\alpha \beta \gamma \xi (\beta^2 + \xi^2) m_Q^2 (m_Q^2 - \mathbf{h}s)
		\Bigg\},
	\end{split}
	\]
	and
	\[
	\begin{split}
		\rho^{\left\langle g^3 G^3\right\rangle } =& -\frac{\left\langle g^3G^3\right\rangle }{2^{16}\pi^8} 
		\int_{\alpha_{\min}}^{\alpha_{\max}} \frac{d\alpha}{\alpha^3} 
		\int_{\beta_{\min}}^{\beta_{\max}} \frac{d\beta}{\beta^3} 
		\int_{\gamma_{\min}}^{\gamma_{\max}} \frac{d\gamma}{\gamma^3} 
		\int_{\xi_{\min}}^{\xi_{\max}} \frac{d\xi}{\xi^3} 
		\frac{\mathbf{h}^3}{(1-\alpha - \beta - \gamma - \xi)^3} 
		\Bigg\{ 
		\Big[(1-\alpha - \beta - \gamma - \xi)^3 + 2\beta^3 + 2\xi^3\Big] \\
		&\Big[4\mathbf{h}^2(m_Q^2 - \mathbf{h}s)^2 
		- 7\mathbf{h}^3 s(m_Q^2 - \mathbf{h}s) 
		+ \mathbf{h}^4 s^2 \Big]
		- \alpha \beta \gamma \xi m_Q^4 \Big[(1-\alpha - \beta - \gamma - \xi)^3 + 12\beta^3 + 12\xi^3\Big]\\
		&-\mathbf{h}\alpha\beta m_Q^2\Big[(1-\alpha - \beta - \gamma - \xi)^3 + 12\beta^3 + 2\xi^3\Big](3m_Q^2 - 4\mathbf{h}s)
		+\mathbf{h}\gamma\xi m_Q^2\Big[(1-\alpha-\beta-\gamma -\xi)^3 + 2\beta^3 + 12\xi^3\Big]\\
		&\times(2m_Q^2 - 3\mathbf{h}s)
		+2\Big[(1-\alpha - \beta - \gamma - \xi)^4 + 2\beta^4 + 2\xi^4\Big]\Big[\mathbf{h}^2 m_Q^2(8m_Q^2-15\mathbf{h}s) 
		+\mathbf{h} m_Q^4(2\gamma\xi-3\alpha\beta) \Big]
		\Bigg\}
	\end{split}
	\]
	for the current \eqref{0},
	\[
	\begin{split}
		\rho^{\text{pert}} = & \frac{3}{2^{14}\pi^8}
		\int_{\alpha_{min}}^{\alpha_{max}} \frac{d\alpha}{\alpha^3} 
		\int_{\beta_{min}}^{\beta_{max}} \frac{d\beta}{\beta^3} 
		\int_{\gamma_{min}}^{\gamma_{max}} \frac{d\gamma}{\gamma^3} 
		\int_{\xi_{min}}^{\xi_{max}} \frac{d\xi}{\xi^3} 
		\frac{\mathbf{h}^3(m_Q^2 - \mathbf{h}s)^3}{(1 - \alpha - \beta - \gamma - \xi)^3} 
		\Bigg\{ 
		\mathbf{h}^2 (m_Q^2 - \mathbf{h}s)^2\\
		& - \left[ 6 \mathbf{h}^3 s + 2 \mathbf{h} m_Q^2 (\alpha \beta + \gamma \xi) \right]  (m_Q^2 - \mathbf{h}s)  
		+ 4 \mathbf{h}^4 s^2 
		+ 4\mathbf{h}^2 s m_Q^2(\alpha \beta + \gamma \xi) 
		+ 4 \alpha \beta \gamma \xi m_Q^4 
		\Bigg\},
	\end{split}
	\]
	\[
	\begin{split}
		\rho^{\left\langle g^2 G^2\right\rangle } =& \frac{m_Q^2\left\langle g^2G^2\right\rangle }{2^{15}\pi^8} 
		\int_{\alpha_{\min}}^{\alpha_{\max}} \frac{d\alpha}{\alpha^3} 
		\int_{\beta_{\min}}^{\beta_{\max}} \frac{d\beta}{\beta^3} 
		\int_{\gamma_{\min}}^{\gamma_{\max}} \frac{d\gamma}{\gamma^3} 
		\int_{\xi_{\min}}^{\xi_{\max}} \frac{d\xi}{\xi^3} 
		\frac{\mathbf{h}^3}{(1-\alpha - \beta - \gamma - \xi)^3} 
		\Bigg\{ 
		\Big[(1-\alpha - \beta - \gamma - \xi)^3 \\
		 &+ 2\beta^3 + 2\xi^3\Big]
		\Big[5\mathbf{h}^2(m_Q^2 - \mathbf{h}s)^2 
		- 12\mathbf{h}^3 s(m_Q^2 - \mathbf{h}s) 
		+ 2\mathbf{h}^4 s^2 
		+ 2\alpha \beta \gamma \xi m_Q^4 
		- 4\mathbf{h} m_Q^2 (\alpha \beta + \gamma \xi)(m_Q^2 - \mathbf{h}s) \\
		&+ 2\mathbf{h}^2 s m_Q^2(\alpha \beta + \gamma \xi)\Big] 
		- 12\mathbf{h} (\gamma \xi^3 + \alpha \beta^3) (m_Q^2 - \mathbf{h}s)^2 
		+ 12(\gamma \xi^3 + \alpha\beta^3) \mathbf{h}^2 s (m_Q^2 - \mathbf{h}s) \\
		&+ 12\alpha \beta \gamma \xi (\beta^2 + \xi^2) m_Q^2 (m_Q^2 - \mathbf{h}s)
		\Bigg\},
	\end{split}
	\]
	and
	\[
	\begin{split}
		\rho^{\left\langle g^3 G^3\right\rangle } =& \frac{\left\langle g^3G^3\right\rangle }{2^{17}\pi^8} 
		\int_{\alpha_{\min}}^{\alpha_{\max}} \frac{d\alpha}{\alpha^3} 
		\int_{\beta_{\min}}^{\beta_{\max}} \frac{d\beta}{\beta^3} 
		\int_{\gamma_{\min}}^{\gamma_{\max}} \frac{d\gamma}{\gamma^3} 
		\int_{\xi_{\min}}^{\xi_{\max}} \frac{d\xi}{\xi^3} 
		\frac{\mathbf{h}^3}{(1-\alpha - \beta - \gamma - \xi)^3} 
		\Bigg\{ 
		\Big[(1-\alpha - \beta - \gamma - \xi)^3 + 2\beta^3 + 2\xi^3\Big] \\
		&\Big[5\mathbf{h}^2(m_Q^2 - \mathbf{h}s)^2 
		- 12\mathbf{h}^3 s(m_Q^2 - \mathbf{h}s) 
		+ 2\mathbf{h}^4 s^2 \Big]
		+ 2\alpha \beta \gamma \xi m_Q^4 \Big[(1-\alpha - \beta - \gamma - \xi)^3 + 12\beta^3 + 12\xi^3\Big]\\
		&-\mathbf{h}\alpha\beta m_Q^2\Big[(1-\alpha - \beta - \gamma - \xi)^3 + 12\beta^3 + 2\xi^3\Big](4m_Q^2 - 6\mathbf{h}s)
		-\mathbf{h}\gamma\xi m_Q^2\Big[(1-\alpha-\beta-\gamma -\xi)^3 + 2\beta^3 + 12\xi^3\Big]\\
		&\times(4m_Q^2 - 6\mathbf{h}s)
		+2\Big[(1-\alpha - \beta - \gamma - \xi)^4 + 2\beta^4 + 2\xi^4\Big]\Big[\mathbf{h}^2 m_Q^2(10m_Q^2-22\mathbf{h}s) 
		+4\mathbf{h} m_Q^4(\gamma\xi+\alpha\beta) \Big]
		\Bigg\}
	\end{split}
	\]
	for the current \eqref{1}, as well as
	\[
	\begin{split}
		\rho^{\text{pert}} = &- \frac{3}{2^{14}\pi^8}
		\int_{\alpha_{min}}^{\alpha_{max}} \frac{d\alpha}{\alpha^3} 
		\int_{\beta_{min}}^{\beta_{max}} \frac{d\beta}{\beta^3} 
		\int_{\gamma_{min}}^{\gamma_{max}} \frac{d\gamma}{\gamma^3} 
		\int_{\xi_{min}}^{\xi_{max}} \frac{d\xi}{\xi^3} 
		\frac{\mathbf{h}^3(m_Q^2 - \mathbf{h}s)^3}{(1 - \alpha - \beta - \gamma - \xi)^3} 
		\Bigg\{ 
		\mathbf{h}^2 (m_Q^2 - \mathbf{h}s)^2\\
		& - \left[ 6 \mathbf{h}^3 s + 2 \mathbf{h} m_Q^2 (\alpha \beta - \gamma \xi) \right]  (m_Q^2 - \mathbf{h}s)  
		+ 4 \mathbf{h}^4 s^2 
		+ 4\mathbf{h}^2 s m_Q^2(\alpha \beta - \gamma \xi) 
		- 4 \alpha \beta \gamma \xi m_Q^4 
		\Bigg\},
	\end{split}
	\]
	\[
	\begin{split}
		\rho^{\left\langle g^2 G^2\right\rangle } =& -\frac{m_Q^2\left\langle g^2G^2\right\rangle }{2^{15}\pi^8} 
		\int_{\alpha_{\min}}^{\alpha_{\max}} \frac{d\alpha}{\alpha^3} 
		\int_{\beta_{\min}}^{\beta_{\max}} \frac{d\beta}{\beta^3} 
		\int_{\gamma_{\min}}^{\gamma_{\max}} \frac{d\gamma}{\gamma^3} 
		\int_{\xi_{\min}}^{\xi_{\max}} \frac{d\xi}{\xi^3} 
		\frac{\mathbf{h}^3}{(1-\alpha - \beta - \gamma - \xi)^3} 
		\Bigg\{ 
		\Big[(1-\alpha - \beta - \gamma - \xi)^3 \\
		& + 2\beta^3 + 2\xi^3\Big] 
		\Big[5\mathbf{h}^2(m_Q^2 - \mathbf{h}s)^2 
		- 12\mathbf{h}^3 s(m_Q^2 - \mathbf{h}s) 
		+ 2\mathbf{h}^4 s^2 
		- 2\alpha \beta \gamma \xi m_Q^4 
		- 4\mathbf{h} m_Q^2 (\alpha \beta - \gamma \xi)(m_Q^2 - \mathbf{h}s)  \\
		&+ 2\mathbf{h}^2 s m_Q^2(\alpha \beta - \gamma \xi)\Big]
		+ 12\mathbf{h} (\gamma \xi^3 - \alpha \beta^3) (m_Q^2 - \mathbf{h}s)^2 
		- 12(\gamma \xi^3 - \alpha\beta^3) \mathbf{h}^2 s (m_Q^2 - \mathbf{h}s) \\
		&- 12\alpha \beta \gamma \xi (\beta^2 + \xi^2) m_Q^2 (m_Q^2 - \mathbf{h}s)
		\Bigg\},
	\end{split}
	\]
	and
	\[
	\begin{split}
		\rho^{\left\langle g^3 G^3\right\rangle } =& -\frac{\left\langle g^3G^3\right\rangle }{2^{17}\pi^8} 
		\int_{\alpha_{\min}}^{\alpha_{\max}} \frac{d\alpha}{\alpha^3} 
		\int_{\beta_{\min}}^{\beta_{\max}} \frac{d\beta}{\beta^3} 
		\int_{\gamma_{\min}}^{\gamma_{\max}} \frac{d\gamma}{\gamma^3} 
		\int_{\xi_{\min}}^{\xi_{\max}} \frac{d\xi}{\xi^3} 
		\frac{\mathbf{h}^3}{(1-\alpha - \beta - \gamma - \xi)^3} 
		\Bigg\{ 
		\Big[(1-\alpha - \beta - \gamma - \xi)^3 + 2\beta^3 + 2\xi^3\Big] \\
		&\Big[5\mathbf{h}^2(m_Q^2 - \mathbf{h}s)^2 
		- 12\mathbf{h}^3 s(m_Q^2 - \mathbf{h}s) 
		+ 2\mathbf{h}^4 s^2 \Big]
		- 2\alpha \beta \gamma \xi m_Q^4 \Big[(1-\alpha - \beta - \gamma - \xi)^3 + 12\beta^3 + 12\xi^3\Big]\\
		&-\mathbf{h}\alpha\beta m_Q^2\Big[(1-\alpha - \beta - \gamma - \xi)^3 + 12\beta^3 + 2\xi^3\Big](4m_Q^2 - 6\mathbf{h}s)
		+\mathbf{h}\gamma\xi m_Q^2\Big[(1-\alpha-\beta-\gamma -\xi)^3 + 2\beta^3 + 12\xi^3\Big]\\
		&\times(4m_Q^2 - 6\mathbf{h}s)
		+2\Big[(1-\alpha - \beta - \gamma - \xi)^4 + 2\beta^4 + 2\xi^4\Big]\Big[\mathbf{h}^2 m_Q^2(10m_Q^2-22\mathbf{h}s) 
		- 4\mathbf{h} m_Q^4(\gamma\xi - \alpha\beta) \Big]
		\Bigg\}
	\end{split}
	\]
	for the current \eqref{2}.
	
	It is defined as \( \mathbf{h} = \dfrac{1}{\frac{1}{\alpha} +\frac{1}{\beta} + \frac{1}{\gamma} + \frac{1}{\xi} + \frac{1}{{1-\alpha-\beta-\gamma-\xi}}} \), and the integration limits of \( \alpha, \beta, \gamma \), and \( \xi \) are given by
	\begin{align*}
		\alpha &= \frac{1}{2} \left[ \left( 1 - \frac{15m_Q^2}{s} \right) \pm \sqrt{\left( 1 - \frac{15m_Q^2}{s} \right)^2 - \frac{4m_Q^2}{s}} \right], \\
		\beta &= \frac{1}{2} \left[ \left( 1 - \alpha - \frac{8\alpha m_Q^2}{\alpha s - m_Q^2} \right) \pm \sqrt{\left( 1 - \alpha - \frac{8\alpha m_Q^2}{\alpha s - m_Q^2} \right)^2 - \frac{4\alpha (1 - \alpha)m_Q^2}{\alpha s - m_Q^2}} \right], \\
		\gamma &= \frac{1}{2} \left[ \left( 1 - \alpha - \beta - \frac{3}{\dfrac{s}{m_Q^2} - \dfrac{1}{\alpha} - \dfrac{1}{\beta}} \right) \pm \sqrt{\left( 1 - \alpha - \beta - \frac{3}{\dfrac{s}{m_Q^2} - \dfrac{1}{\alpha} - \dfrac{1}{\beta}} \right)^2 - 4\frac{1-\alpha-\beta}{\dfrac{s}{m_Q^2} - \dfrac{1}{\alpha} - \dfrac{1}{\beta}}} \right], \\
	\end{align*}
	and
	\begin{align*}
		\xi &= \frac{1}{2} \left[ (1 - \alpha - \beta - \gamma) \pm \sqrt{(1 - \alpha - \beta - \gamma)^2 - 4\frac{1 - \alpha - \beta - \gamma}{\dfrac{s}{m_Q^2} - \dfrac{1}{\alpha} - \dfrac{1}{\beta} - \dfrac{1}{\gamma}}} \right].
	\end{align*}	
	
\end{widetext}

\section{NUMERICAL ANALYSIS AND DISCUSSIONS}\label{three}
In QCD sum rules, one expands the correlation function by perturbation and non perturbation approximation in the OPE caculation, and introduces a complex dispersion integral structure including hadronic resonance and continuous spectrum at the phenomenological level.
In order to reliably extract the information of hadronic resonance states, it is necessary to determine the appropriate working windows of the continuum threshold $\sqrt{s_0}$ and the Borel parameter $M^2$.
The threshold $\sqrt{s_0}$ represents the starting energy point of the continuum state, and the energy gap $\sqrt{s_0} - M_H$ is typically estimated at approximately 0.3 – 0.8 GeV \cite{RN46}.
The Borel parameter is determined from two considerations. 
First, the lower bound is determined by the convergence requirement of the OPE, ensuring that higher-order term contributions can be safely neglected. 
Second, the upper bound is constrained by the condition that pole contribution should significantly exceed continuum contribution. 

The input parameters are taken as $\langle g^2 G^2 \rangle$ = 0.88 $\pm$ 0.25  $\text{GeV}^4$, $\langle g^3 G^3 \rangle $= 0.58 $\pm$ 0.18  $\text{GeV}^6$ \cite{RN25,RN26,RN46,RN47,RN48,RN49,RN50} and $m_Q$ is set to the running charm mass $m_c $= 1.2730 $\pm$ 0.0046 $\text{GeV}$ \cite{RN51}.
For the $cccc\bar{c}$ represented by the current \eqref{0}, based on the typical energy gap between the continuum spectrum and the resonance state, $\sqrt{s_0}$ takes the value in the range of $8.0-8.4$ GeV.
The lower limit of the Borel parameter is taken to be $M^2 \geq 3.5\,\text{GeV}^2$. 
The pole relative contribution at  $\sqrt{s_0}$ = 8.2 GeV reaches a critical value of 50\% at $M^2$ = 4.1 $\text{GeV}^2$, and this contribution decreases monotonically with increasing $M^2$.
Based on the pole dominance condition, the choice of the Borel window at $\sqrt{s_0}$ = 8.2 GeV is finalized as $M^2 \in [3.5, 4.1]$ $\text{GeV}^2$.
By analyzing the pole dominance of the system and the OPE convergence test, it can be obtained that when $\sqrt{s_0} = 8.0 $ GeV, the corresponding Borel window is $M^2 \in [3.5, 3.6]$ $\text{GeV}^2$; while when $\sqrt{s_0}$ increases to 8.4 GeV, the Borel window extends to $M^2 \in [3.5, 4.5]$ $\text{GeV}^2$ accordingly.

\begin{figure}[htbp]
	\centerline{\epsfysize=7.18truecm
		\epsfbox{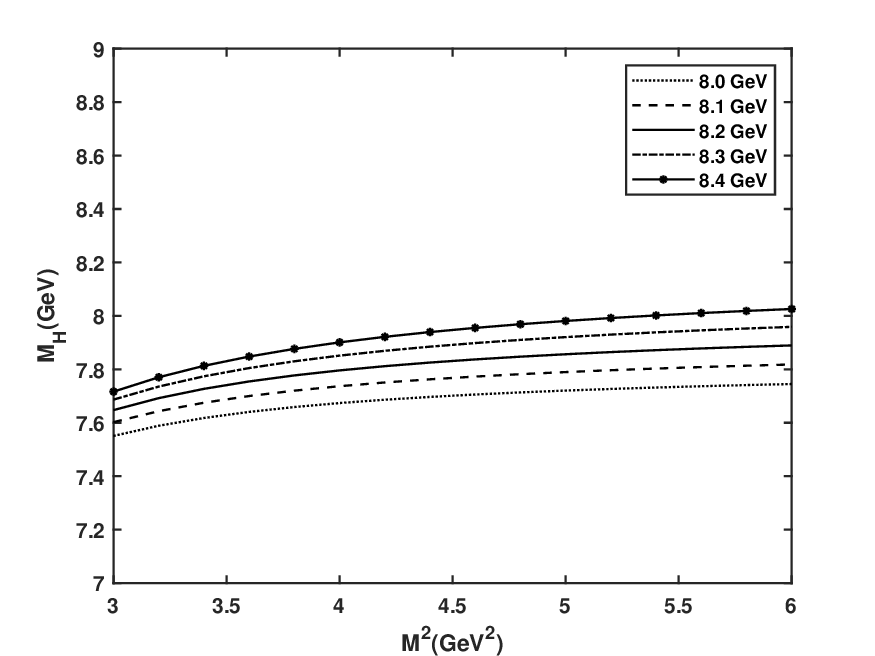}}
	\caption{The Borel Curves for $cccc\bar{c}$ state of current \eqref{0}. 
		The Borel windows of $M^2$ are $3.5–3.6$ $\text{GeV}^2$ for $\sqrt{s_0}$ = 8.0 GeV, $3.5–4.1$ $\text{GeV}^2$ for $\sqrt{s_0}$ = 8.2 GeV, and $3.5–4.5$ $\text{GeV}^2$ for $\sqrt{s_0}$ = 8.4 GeV, respectively.}
	\label{c}
\end{figure}

\begin{figure}[htbp]
	\centerline{\epsfysize=7.18truecm
		\epsfbox{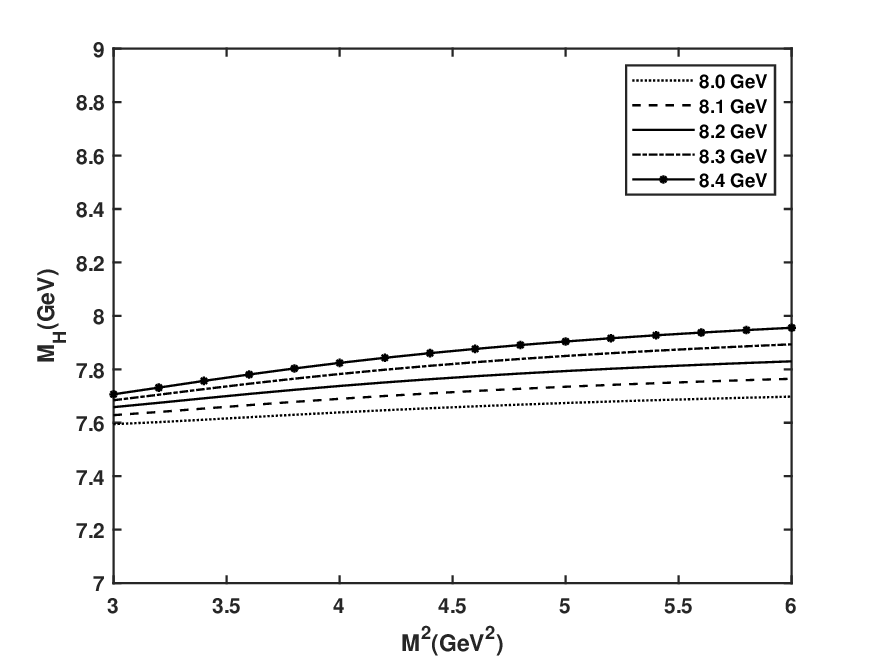}}
	\caption{The Borel Curves for $cccc\bar{c}$ state of current \eqref{1}.
		The Borel windows of $M^2$ are $3.5–3.7$ $\text{GeV}^2$ for $\sqrt{s_0}$ = 8.0 GeV, $3.5–4.4$ $\text{GeV}^2$ for $\sqrt{s_0}$ = 8.2 GeV, and $3.5–5.1$ $\text{GeV}^2$ for $\sqrt{s_0}$ = 8.4 GeV, respectively.}
	\label{cc}
\end{figure}

\begin{figure}[htbp]
	\centerline{\epsfysize=7.18truecm
		\epsfbox{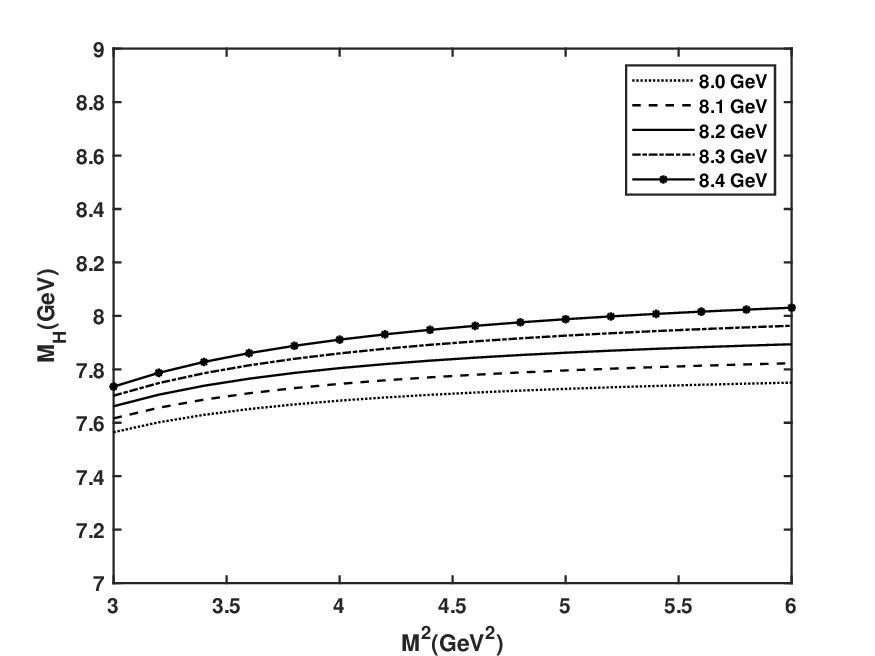}}
	\caption{The Borel Curves for $cccc\bar{c}$ state of current \eqref{2}.
		The Borel windows of $M^2$ are $3.5–3.6$ $\text{GeV}^2$ for $\sqrt{s_0}$ = 8.0 GeV, $3.5–3.9$ $\text{GeV}^2$ for $\sqrt{s_0}$ = 8.2 GeV, and $3.5–4.5$ $\text{GeV}^2$ for $\sqrt{s_0}$ = 8.4 GeV, respectively.}
	\label{ccc}
\end{figure}

In the calculations, the parameters are kept at their center values.
The Borel curves for the  $cccc\bar{c}$ pentaquark states are shown in Fig. \ref{c}, Fig. \ref{cc}, and Fig. \ref{ccc}.
The extracted masses of bound states within the optimally chosen working windows are $M_{H1} =7.79^{+0.16}_{-0.16}$ $\text{GeV}$ for current \eqref{0},  $M_{H2} =7.76^{+0.18}_{-0.15}$ $\text{GeV}$ for current \eqref{1}, and  $M_{H3} =7.81^{+0.15}_{-0.17}$ $\text{GeV}$ for current \eqref{2}.
After systematically accounting for uncertainties in the input parameters, the extracted masses of the $cccc\bar{c}$  pentaquark states are $M_{H1} = 7.79^{+0.16+0.02}_{-0.16-0.01}$ $\text{GeV}$, $M_{H2} = 7.76^{+0.18 +0.05}_{-0.15-0.03}$ $\text{GeV}$, as well as $M_{H3} = 7.81^{+0.15+0.02}_{-0.17-0.01}$ $\text{GeV}$.
The first uncertainty arises from the variation in the value of the continuous threshold parameter $\sqrt{s_0}$ and the Borel parameter $M^2$, while the second arises from the errors in the QCD parameters.
These results can also be concisely expressed as $M_{H1} = 7.79^{+0.18}_{-0.17}$ $\text{GeV}$, $M_{H2} = 7.76^{+0.23}_{-0.18}$ $\text{GeV}$, and $M_{H3} = 7.81^{+0.17}_{-0.18}$ $\text{GeV}$.

\begin{figure}[htbp]
	\centerline{\epsfysize=7.18truecm
		\epsfbox{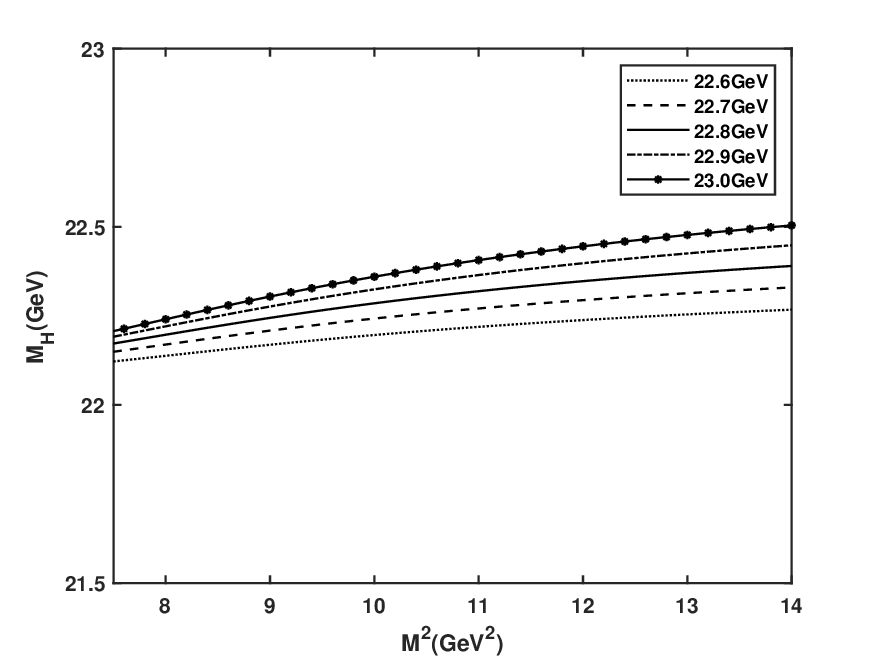}}
	\caption{The Borel Curves for $bbbb\bar{b}$ state of current \eqref{0}.
		The Borel windows of $M^2$ are $ 8–10.6$ $\text{GeV}^2$ for $\sqrt{s_0}$ = 22.6 GeV, $8.0-12.3$ $\text{GeV}^2$ for $\sqrt{s_0}$ = 22.8 GeV, and $8.0-13.7$ $\text{GeV}^2$ for $\sqrt{s_0}$ = 23.0 GeV, respectively.}
	\label{b}
\end{figure}

\begin{figure}[htbp]
	\centerline{\epsfysize=7.18truecm
		\epsfbox{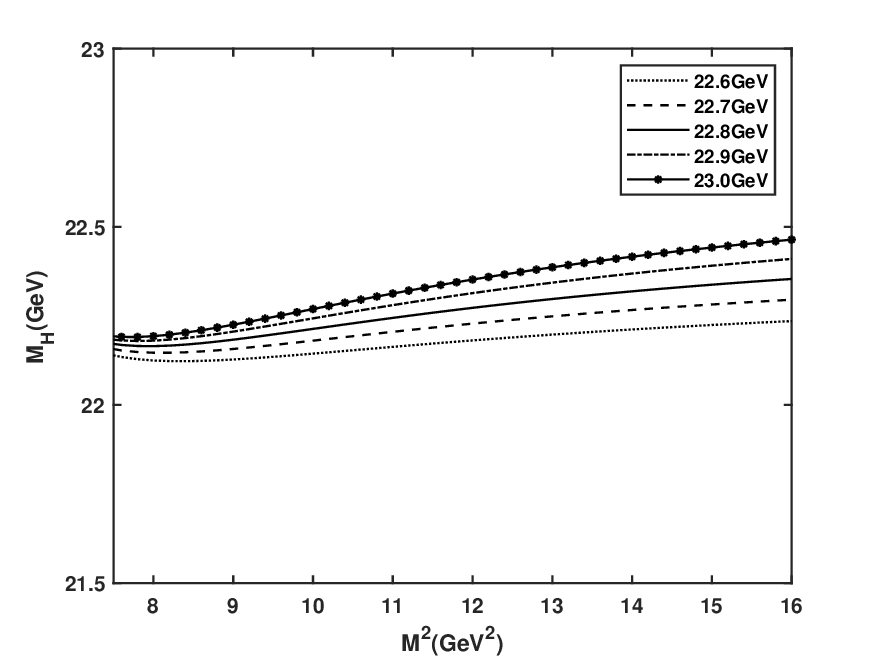}}
	\caption{The Borel Curves for $bbbb\bar{b}$ state of current \eqref{1}.
		The Borel windows of $M^2$ are $ 8–12.6$ $\text{GeV}^2$ for $\sqrt{s_0}$ = 22.6 GeV, $8.0-14.3$ $\text{GeV}^2$ for $\sqrt{s_0}$ = 22.8 GeV, and $8.0-16.0$ $\text{GeV}^2$ for $\sqrt{s_0}$ = 23.0 GeV, respectively.}
	\label{bb}
\end{figure}

\begin{figure}[htbp]
	\centerline{\epsfysize=7.18truecm
		\epsfbox{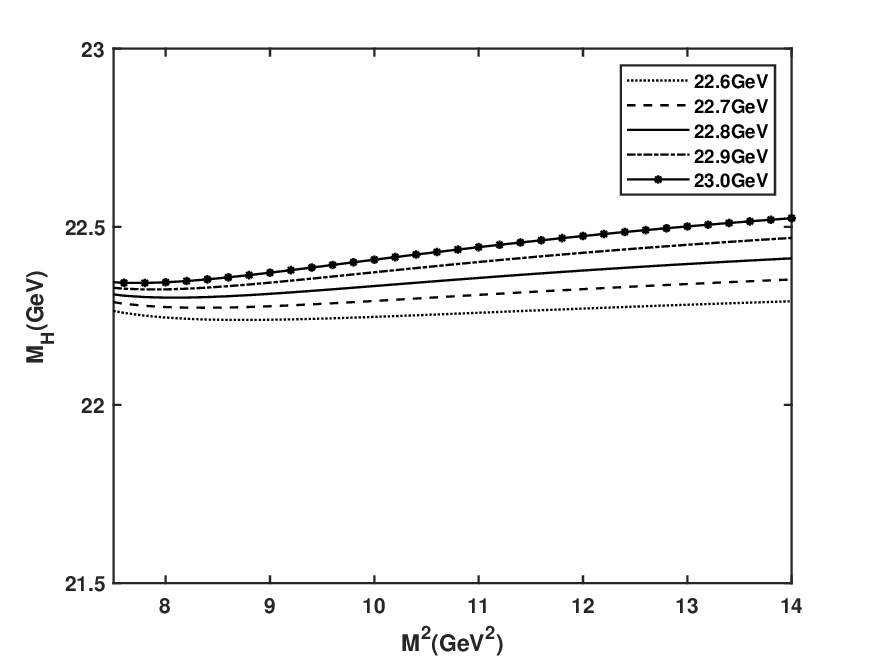}}
	\caption{The Borel Curves for $bbbb\bar{b}$ state of current \eqref{2}.
		The Borel windows of $M^2$ are $ 8–10.6$ $\text{GeV}^2$ for $\sqrt{s_0}$ = 22.6 GeV, $8.0-12.2$ $\text{GeV}^2$ for $\sqrt{s_0}$ = 22.8 GeV, and $8.0-13.7$ $\text{GeV}^2$ for $\sqrt{s_0}$ = 23.0 GeV, respectively.}
	\label{bbb}
\end{figure}

For the systematic analysis of the fully bottomed $bbbb\bar{b}$ pentaquark states, we use the running bottom quark mass $m_b = 4.183\pm0.007$ GeV \cite{RN51} instead of the mass $m_c$.
The threshold $\sqrt{s_0}$ range for the $bbbb\bar{b}$ state is taken as $22.6-23.0$ GeV, and the Borel curves are shown in Fig. \ref{b}, Fig. \ref{bb}, and Fig. \ref{bbb}.
The final calculated masses of the $bbbb\bar{b}$ states are $M_{H1} = 22.35^{+0.17 +0.02}_{-0.11 -0.08}$ GeV, $M_{H2} = 22.27^{+0.20 +0.02}_{-0.15-0.07}$ GeV, and $M_{H3} = 22.37^{+0.15+0.03}_{-0.13-0.07}$ GeV.
These results can also be briefly written as $M_{H1} = 22.35\pm0.19$ $\text{GeV}$, $M_{H2} = 22.27\pm0.22$ $\text{GeV}$, and $M_{H3} = 22.37^{+0.18}_{-0.20}$ $\text{GeV}$. 

In Table \ref{table:1}, we collect our final results for the mass spectra of fully heavy pentaquark states and make a comparison with other works.
By comparison, the masses of ${\frac{3}{2}}^-$ pentaquarks obtained in our works are smaller than the rest of results shown, which are gained from various theoretical models, such as the CMI model, chiral quark model and quark delocalization color screening model, constituent quark model, lattice-QCD inspired quark model, MIT bag model, extended Gursey-Radicati formalism, effective quark mass and screened charge scheme etc. 
The main reason causing the mass differences is due to different theoretical treatments on sophisticated nonperturbative QCD effects. 
Unlike the model-dependent approaches, the QCD sum rule is one nonperturbative method firmly founded on the basic QCD theory. 
Our results are consistent with the QCD sum rule results of Ref. \cite{RN16} within the errors (there some slight differences come from our inclusion of high dimension three-gluon condensates and choice of different Borel windows), which are comparatively smaller than other theoretical models. 

Moreover, we note that Ref. \cite{RN17} states that there is no compact structure between the baryon and the meson about the ${\frac{5}{2}}^-$ state using diffusion Monte Carlo calculation. 
In this work, we devote to systematically study and present the spectra of fully heavy pentaquark states from QCD sum rules, in expectation of that they could exist and would be found in future experiments. 
For the sake of comparison, we list the constituent baryon and meson masses of fully heavy pentaquark states in Table \ref{table:2}.
Seeing from that the extracted mass values with QCD sum rules are lower than thresholds of the baryon and meson constituents, one expects that they could form some loosely bound states.

\begin{table*}[htbp]\caption{The mass spectra of fully heavy pentaquark states (mass in unit of MeV except for ``Our works").
		Among them, CMI is CMI model, ChQM-QDCSM is chiral quark model and quark delocalization color screening model, CQM is constituent quark model, LQCD-IQM is lattice-QCD inspired quark model, MIT is MIT bag model, QCDSR is QCD sum rules, EGRF is extended Gursey-Radicati formalism, and EQM is effective quark mass and screened charge scheme.}
	\centering
	{\begin{tabular}{*{11}{c}} \hline\hline
			State  & $J^{P}$ &  Ours (GeV)  & CMI\cite{RN8} &  ChQM-QDCSM\cite{RN10}  & CQM\cite{RN11} & LQCD-IQM\cite{RN12} &  MIT\cite{RN13} & QCDSR  \cite{RN16} &  EGRF\cite{RN20} & EQM\cite{RNad12} \\ \hline 
			$P_{cccc\bar{c}}$ &  $\frac{3}{2}^{-}$ & $7.41^{+0.27}_{-0.31}$ \cite{RNzjr} & 7864 &   & 8144.6 & 8095 & 8229  & $7628\pm112$ & $8425.85\pm91.12$ & 8547.4  \\
			
			& $\frac{3}{2}^{+}$  &  $7.79^{+0.18}_{-0.17}$  \\
			
			& $\frac{5}{2}^{-}$  &  $7.76^{+0.23}_{-0.18}$ & & & & & & &    $8540.85\pm91.61$ & 8562.5 \\
			
			& $\frac{5}{2}^{+}$  &  $7.81^{+0.17}_{-0.18}$  \\	    
			
			$P_{bbbb\bar{b}}$ & $\frac{3}{2}^{-}$  &  $21.60^{+0.73}_{-0.22}$ \cite{RNzjr} & 23775 & 23748.2-23752.3 & 24210.9 & 24035 & 24761  & $21982\pm144$ &   $25752.90\pm172.06$ & 25214.8 \\       
			
			& $\frac{3}{2}^{+}$  &  $22.35\pm0.19$\\
			
			& $\frac{5}{2}^{-}$  &  $22.27\pm 0.22$ & & & & & & &    $25867.90\pm172.32$ & 25216.3\\
			
			& $\frac{5}{2}^{+}$  & $22.37^{+0.18}_{-0.20}$   \\    
			\hline\hline
	\end{tabular}}
	\label{table:1}
\end{table*}

\begin{table}[htbp]\caption{The masses of mesons and baryons}
	\centering
	{\begin{tabular}{ p{1.5cm} p{3cm} cp{.0cm}} \hline\hline
			State &   & $M(\text{MeV})$\\ \hline
			
			Mesons &  &     \\
			
			$\eta_c(1S)$  &   &  $2984.1\pm 0.4$\cite{RN51}\\
			
			$\eta_b(1S)$  &   &  $9398.7\pm 2.0$\cite{RN51}\\
			
			$J/\psi(1S)$  &  &$3096.900\pm0.006$ \cite{RN51}\\
			
			$\Upsilon(1S)$ &  &  $9460.40\pm0.10$\cite{RN51} \\
			
			$\chi_{c0}(1P)$ &  &  $3414.71\pm0.30$ \cite{RN51} \\
			
			$\chi_{c1}(1P)$ &  &  $3510.67\pm0.05$ \cite{RN51} \\	    
			
			$\chi_{b0}(1P)$ &  &  $9859.44\pm0.42\pm0.31$\cite{RN51}\\
			
			$\chi_{b1}(1P)$ &  &  $9892.78\pm0.26\pm0.31$\cite{RN51}\\
			Baryons & &   \\       
			
			$\Omega_{ccc}$ & &  $4796$\cite{RNad13} \\
			
			$\Omega_{bbb}$&  & $14366$\cite{RNad13}   \\    
			\hline\hline
	\end{tabular}}
	\label{table:2}
\end{table}

\section{SUMMARY}
In the framework of QCD sum rule method, we have systematically studied fully heavy pentaquark $QQQQ\bar{Q}$ states and finally obtained their mass spectra.
For three different configurations of currents, the corresponding masses are presented as $7.79^{+0.18}_{-0.17}$ GeV, $7.76^{+0.23}_{-0.18}$ Gev, $7.81^{+0.17}_{-0.18}$ GeV for the $cccc\bar{c}$ states, and $22.35\pm0.19$ GeV,  $22.27\pm+0.22$ GeV,
$22.37^{+0.18}_{-0.20}$ GeV for the $bbbb\bar{b}$ states, respectively. 
These fully heavy pentaquark states might be experimentally searched for in strong decay channels, provided that they allow strong decay processes to occur.
Among them, the states with $J^P = \frac{3}{2}^{+}$ could be detected in the $\Omega_{QQQ} \chi_{Q0}$ invariant mass spectrum, the states  with $J^P = \frac{5}{2}^{-}$ in the $\Omega_{ccc} J/\psi$ or $\Omega_{bbb} \Upsilon$ invariant mass spectrum, and the states with $J^P = \frac{5}{2}^{+}$ in the $\Omega_{QQQ} \chi_{Q1}$ invariant mass spectrum, respectively.
Anyhow, subsequent theoretical studies and intensive experimental observations will provide an important basis for revealing the nature of these peculiar fully heavy pentaquark states.
%%%%%%%%%%%%%%%%%%%%%%%%%%%%%%%%%%%%%%%%%%%%%%%%%%%%%%%%%%%%%%%%%%%

\begin{acknowledgments}
	This work was supported by the National Natural Science Foundation of China under Contract No. $11475258$. The authors are very grateful to Xiang Liu for recent communication and helpful discussion.
\end{acknowledgments}

%%%%%%%%%%%%%%%%%%%%%%%%%%%%%%%%%%%%%%

\end{document}